\documentclass[aps,prl,twocolumn,nofootinbib]{revtex4-2}

\usepackage{amsmath}
\usepackage{amssymb}
\usepackage{bm}
\usepackage[T1]{fontenc}
\usepackage{lmodern}
\usepackage[colorlinks=true,allcolors=blue]{hyperref}
\usepackage{xurl}

\begin{document}

\title{What time is it in Assembly Space? Cronin and Walker's ``The Physics of Causation'' fails elementary dimensional analysis (Comment on arXiv:2601.00515)
}

\author{Philip Moriarty}
\email{philip.moriarty@nottingham.ac.uk}
\affiliation{School of Physics and Astronomy, University of Nottingham, Nottingham NG7 2RD, United Kingdom}

\date{\today}

\begin{abstract}
Cronin and Walker claim that causation is quantified by an ``assembly index'', a count of the steps needed to build an object. Remarkably, they equate ``continuous time (as measured by clocks)'' to a multiple of the index, which is a dimensionless number. Their argument treats one parameter as simultaneously a rate, a time interval, and a pure number, and provides no timescale to convert index to clock time. This dimensional incoherence betrays, at best, an ill-conceived and ill-formed theory, and fatally undermines two central claims. First, the exponential weighting at the heart of what they call the assembly equation can be obtained only by supplying a timescale, which they never do, or by recasting clock time as a dimensionless count of assembly steps. In the latter case, their ``derivation'' trivially reduces to Bernoulli's seventeenth-century compound-interest limit. Second, lacking rates or energy scales, their account cannot predict how quickly, or under what conditions, anything forms, and yet formation ``in finite time'' is their criterion for distinguishing spontaneous from ``selected'' objects. Assembly theory is thus incapable of addressing causation.
\end{abstract}

\maketitle

Cronin, Walker (CW) and co-workers have promoted ``assembly theory'' (AT) as a means of quantifying how much ``selection'' is needed to produce a given object, with a primary aim of detecting life without presupposing its chemistry~\cite{Marshall2021,Sharma2023}. They have asserted that AT explains evolution and quantifies causation~\cite{Sharma2023} and, most recently, that it provides ``the foundation for a theory of physics''~\cite{CroninWalker2026}. These inflated claims have drawn extensive, sustained, and well-founded criticism~\cite{Uthamacumaran2024,Abrahao2024,Ozelim2026,Hazen2024,Jaeger2024,Lynch2025,Myers2023}.

AT's core quantity is an object's ``assembly index'', $a$, the minimum number of pairwise joining operations needed to construct it from basic building blocks, reusing previously built sub-objects. Objects with a high assembly index found at high copy number are, CW argue, the products of selection~\cite{Marshall2021,Sharma2023}. ``Selection'' here is not meant in the Darwinian sense but as ``the act of reducing a set of options to only those that persist over time''~\cite{CroninWalker2026}. This definition is inherently temporal, and CW explicitly invoke ``continuous time (as measured by clocks)''~\cite{CroninWalker2026}. Time measured by clocks necessarily has dimensions $[\mathrm{T}]$.

That the assembly index is nothing more than a compression-based complexity measure, of a kind introduced half a century or more ago~\cite{Kolmogorov1965,LZ1976}, has been argued in detail by Zenil and co-workers~\cite{Uthamacumaran2024,Abrahao2024,Ozelim2026} (see also Ref.~\cite{Masierak2026}). I will not revisit those compelling critiques here, but focus instead on CW's central physical claim: that AT accounts for causation.

CW declare causation to be a ``material property'' and the assembly index ``a newly identified intrinsic physical invariant'', placing causation on a par with ``charge, spin, or other quantum numbers, or the speed of light $c$''~\cite{CroninWalker2026,DimensionsNote}. Physical quantities, however, are subject to a basic requirement that no amount of reinterpretation can circumvent: dimensional consistency. As I show below, CW's treatment of time does not withstand even elementary dimensional analysis (of the type taught in first-year courses across many disciplines, not least physics)~\cite{Robinett2025}.

CW's only explicit link between the assembly index and physical time is a short ``derivation'' of the exponential weighting in their ``assembly equation''~\cite{CroninWalker2026}. They assume that a ``virtual copy number'', $v$, grows as
\begin{equation}
v(t)=(1+h)^{t},
\label{eq:v}
\end{equation}
where $h$ is ``the growth rate'' and $t$ is ``a continuous time over which objects were created''. Equation~(\ref{eq:v}) is physically meaningless. Its left-hand side, a copy number, is dimensionless, yet with $h$ a growth rate and $t$ a time, the right-hand side adds a pure number to a rate and raises the result to the power of a time. Dimensionally sound alternatives, e.g., $v=e^{ht}$ or $v=(1+h\tau)^{t/\tau}$ for some period $\tau$, each require a timescale. CW supply none.

Their next step is to write
\begin{equation}
\Delta t = h = 1/P ,
\label{eq:h}
\end{equation}
with $P$ ``the number of possibilities from which a single virtual object is selected'', explaining that ``the growth rate of virtual copies in the past is defined as the minimum unit of time to form a single (virtual) object''~\cite{CroninWalker2026}. Equation~(\ref{eq:h}) assigns $h$ three incompatible identities in one line: a growth rate, with dimensions $[\mathrm{T}^{-1}]$; a time interval $\Delta t$, with dimensions $[\mathrm{T}]$; and the reciprocal of a count of possibilities, which is dimensionless. Equation~(\ref{eq:h}) is therefore also physically meaningless. 

From this, CW relate the assembly index to time by
\begin{equation}
a = t\cdot\Delta t = t/P ,
\label{eq:a}
\end{equation}
which is described in Ref.~\cite{CroninWalker2026} as ``the connection between causal time (measured in assembly steps, $a$) and continuous time (as measured by clocks, $t$)'', so that the ``total coordinate time'' is $t=aP$. The dimensions fail whichever identity of $h$ in Eq.~(\ref{eq:h}) is chosen. If $h$ is a time interval, as both ``minimum unit of time'' and the symbol $\Delta t$ strongly suggest, then $t\cdot\Delta t$ has dimensions $[\mathrm{T}^2]$, not those of the pure number $a$. If $h$ is a rate, Eq.~(\ref{eq:v}) fails as described above, and Eq.~(\ref{eq:h}) equates that rate to a pure number. If $h$ is dimensionless, as $h=1/P$ requires, then $t$ must be too, contradicting CW's statement that $t$ is time ``as measured by clocks'', and $t=aP$ equates a time to a product of pure numbers. No assignment of dimensions to $h$ satisfies both Eqs.~(\ref{eq:v})--(\ref{eq:a}) and CW's definitions.

CW go on to substitute $t=aP$ into Eq.~(\ref{eq:v}) to obtain \mbox{$v(a)=(1+1/P)^{aP}$}, and take $P\to\infty$ to yield $v(a)=e^{a}$. This exponential is the weighting factor in what they term the ``Assembly'' $A$ of an ensemble of objects [Eq.~(3) of Ref.~\cite{CroninWalker2026}]: \begin{equation} A = \Omega\frac{1}{N_T}\sum_{i,\,n_i>1} n_i\, e^{a_i}. \label{eq:A} \end{equation} Here the sum runs over those distinguishable object types $i$ in the ensemble that occur more than once; $a_i$ is the assembly index of type $i$ and $n_i$ its copy number; $N_T$ is the total number of objects of all types; and $\Omega$ is a constant ``measured in kJ\,mol$^{-1}$''~\cite{CroninWalker2026}. They describe $A$ variously as the ``total causation'' of the ensemble and as ``a causal density per observed object''.

Equation~(\ref{eq:A}) differs from the definition of Assembly given earlier by CW and co-workers~\cite{Sharma2023}, \textit{viz.}
\begin{equation}
A = \frac{1}{N_T}\sum_{i=1}^{N} (n_i-1)\,e^{a_i},
\label{eq:A2023}
\end{equation}
where $N$ is the number of distinct object types. This $A$ is dimensionless. Replacing $n_i-1$ by $n_i$ (with \mbox{$n_i>1$}) matters little at high copy number. The substantive change is the introduction of $\Omega$ in Eq.~(\ref{eq:A}), which Ref.~\cite{CroninWalker2026} makes without comment.

Every factor in Eq.~(\ref{eq:A}) other than $\Omega$ is a dimensionless number, so CW's ``total causation'' now has the dimensions of a molar energy, such as an activation energy. However, energy plays no part in CW's analysis beyond a cursory comparison of heats of formation for a single pair of isomers. Despite giving $A$ the dimensions of an energy, $[\mathrm{ML^2T^{-2}}]$, per mole, CW state that it plays ``an analogous role to entropy'', and plot it alongside a thermodynamic entropy of formation, on twin axes in kJ\,mol$^{-1}$ and kJ\,K$^{-1}$\,mol$^{-1}$ respectively (Fig. 8 of Ref.~\cite{CroninWalker2026}). Thermodynamic entropy has dimensions $[\mathrm{ML^2T^{-2}\Theta^{-1}}]$ and can thus be related to $A$ only through a temperature. None appears anywhere in AT~\cite{EntropyNote}. In addition, $\Omega$ takes its units only ``for molecules'': the dimensions of causation would thus depend on the kind of object, which is incompatible with CW's claim that causation is ``a feature like charge''. Moreover, CW arbitrarily set $\Omega$ to unity.

Rescuing the algebra necessitates abandoning the claim it was meant to support: a connection between the assembly index and clock time. Equations~(\ref{eq:v})--(\ref{eq:a}) are consistent only if $t$ is not time ``as measured by clocks'' but a dimensionless count of ticks, contrary to CW's explicit definition. Their equations then reduce to the compound-interest formula at a nominal rate of 100\% per assembly step, compounded $P$ times per step, and the limit $P\to\infty$ is Bernoulli's seventeenth-century route to $e$~\cite{Maor1994}. The periodic rate $h$ and the interval $\Delta t$ coincide in Eq.\ref{eq:h} as $1/P$ only because both the nominal rate and the step duration $T_0$ have implictly been set to unity. Restoring them gives $v=e^{ra}$ for a rate $r$ per step, and clock time $t=aT_0$. However, nothing in CW's account establishes $r$ or $T_0$. Either $t$ is clock time, in which case Eqs.~(\ref{eq:v})--(\ref{eq:a}) are physically meaningless and, since $t=aP$, the limit $P\to\infty$ implies that every object takes infinite time to form; or it is not, and the weighting $e^{a}$ rests on the implicit, unexplained choice that $r=1$. Either way, the ``derivation'' is, to use Pauli's acerbic assessment of work so ill-defined as to be meaningless, ``not even wrong''~\cite{Peierls1960}.

Without rates or timescales, CW's distinction between spontaneous and selected objects is untestable. Objects below their threshold are said to arise spontaneously ``in finite time'', whereas those above it ``cannot form in abundance in finite time without being promoted by selection''~\cite{CroninWalker2026}. ``Finite time'' is defined nowhere, and the threshold depends only on the number of objects, a branching factor and a detection limit, not on time, temperature, energy, or rate. CW dismiss ``unstructured (or random) biases in reaction rates'' as insufficient to produce objects above the threshold, yet never quantify a single rate. Any process with a constant non-zero rate $\nu$ has a finite mean waiting time; what separates the spontaneous from the improbable is a timescale~\cite{AdamsNote}.

CW might counter that AT is not meant to reduce to kinetics: they decline to weight assembly pathways by probability because doing so introduces ``additional environmental assumptions''~\cite{CroninWalker2026}. But those ``assumptions'' are the temperatures, free energies, barriers, and rates that select among pathways. Rates can invert AT's ordering: a high-index object reached via low barriers can form faster, and more abundantly, than a low-index counterpart facing a single high barrier. In setting kinetics and thermodynamics aside, CW discard the only information distinguishing the pathways actually followed from those merely possible, despite conceptualising objects ``as entities defined by their possible formation histories''~\cite{Sharma2023}.

Finally, a theory claimed to extend established physics must reduce to it in the appropriate limits, as special relativity does to Newtonian mechanics for speeds $u\ll c$, and quantum mechanics to classical mechanics through the correspondence principle. AT has no similar ``kinetic limit''~\cite{ReductionNote}. It cannot begin to express, let alone recover, the rate laws that govern how molecules form. Nor can it dispense with rates while claiming to account for ``continuous time (as measured by clocks)''. Causation, the property claimed to distinguish AT from other complexity measures, therefore lies entirely beyond its reach.

\noindent \textbf{Acknowledgements.} My thanks to colleagues, in a variety of disciplines, who read and commented on a draft of this manuscript. Anthropic's Claude (Opus 5.5) was used iteratively to cross-check quotations and characterisations against CW's text (and other AT-related publications), to formulate the strongest possible case for AT (i.e. to explore the most charitable interpretations of CW's arguments), and to copy-edit successive drafts. The arguments, conclusions, and any remaining errors are my own.


\begin{thebibliography}{99}
\bibitem{DimensionsNote}
CW's list places dimensionless quantum numbers alongside a dimensional constant, $c$ $[\mathrm{LT^{-1}}]$. Charge $[\mathrm{IT}]$ and spin $[\mathrm{ML^2T^{-1}}]$ yield quantum numbers only on division by $e$ and $\hbar$. A dimensionless count is unobjectionable in itself; what AT lacks is an analogue of $e$ or $\hbar$ through which its counts could be equated with dimensional quantities, such as clock time.
\bibitem{ReductionNote}
Fundamentally, AT cannot be reconciled with statistical mechanics. Its claim that high-index objects will not form in abundance without selection, independent of energy and temperature, contradicts the kinetics and thermodynamics that underpin all of chemistry, in which free energies and rates determine whether ordered structures self-assemble spontaneously (and in vast numbers). AT contains no parameter whose limit could recover thermodynamics or kinetics, as, for example, $u/c\to0$ recovers Newtonian mechanics. Restoring the energies, temperatures, barriers, and rates that consistency demands would not generalise AT but leave nothing of it.
\end{thebibliography}
\end{document}